\documentstyle[aps,epsf,amsmath,amssymb]{revtex}

\newcommand{\bra}[1]{\mathop{\langle\left.{#1}\right|}\nolimits}
\newcommand{\ket}[1]{\mathop{\left|{#1}\right.\rangle}\nolimits}

\begin{document}

\title{The Information Capacity of the $\Lambda$-System--Photon Field Channel}
\author{\bf B. A. Grishanin and V. N. Zadkov\\ {\em Physics
Faculty and International Laser Center}\\ {\em M. V. Lomonosov Moscow State
University, Moscow, 119899 Russia}\\ \small e-mail:
grishan@comsim1.phys.msu.su \\ \small e-mail: zadkov@comsim1.phys.msu.su}
\maketitle

\vspace{0.03\textheight}
\begin{abstract}{\bf Abstract}---The potentially attainable information
capacity of a radiatively stable $\Lambda$-system, viewed as the input of a
quantum information channel, is studied. The output of the channel is formed
by the states of the photon field with optical excitation frequencies, which
are created as a result of Raman pumping. The analysis is based on the
notion of coherent information.
\end{abstract}

\section{Introduction}\label{section:intro}

For the experiments in the newest fields of physics--quantum computing,
quantum commu\-ni\-ca\-tion, and quantum cryptography \cite{1,2} the
quantitative analysis of the possibly available amount of the specifically
quantum coherent information \cite{3,4} is of substantial interest, since
this parameter determines the potential information content of the obtained
data. We have previously shown \cite{5} how the general definition of the
coherent information can be applied to the analysis of various physical
models of quantum channels, which in general may have qualitatively
different structure of the input and output state spaces.

In this work, we consider another important example of a quantum channel:
the $\Lambda$-system \cite{6} interacting with a free-space photon field.
Atomic $\Lambda$-systems are particularly interesting for experimental
implementations of the quantum-computing operations since the quantum
information in this system can be stored in radiatively stable two-level
subsystems of separate atoms and manipulated via radiatively excited states
with the help of laser radiation \cite{7}. The specific feature of the
channel considered in this paper is the absence of any information at the
output in the absence of laser excitation, which would otherwise act on the
input states and lead to the excitation of radiatively active states. Thus,
one can distinguish a special class of active channels whose capacity is
intrinsically mined by the external disturbance. The examples of quantum
channels considered in \cite{5} include the following channels belonging
this class: (i) a channel between allowed transition and a forbidden
transition of a (hydrogen) atom that are coupled by an applied external
electric field and (ii) a channel between a pair of two- level atoms (TLAs)
in the presence of an external excitation that entangles their states. The
remaining examples belong to the class of passive channels, in which the
information exchange takes place in the absence of an external perturbation
as well.

\section{The definition and the physical meaning of coherent
information}\label{section:def}

Consider a quantum channel whose input state is described by a density
matrix $\hat\rho_{\rm in}$ and that corresponds to a superoperator
transformation ${\cal C}$, which relates the input and output density
matrices, $\hat\rho_{\rm out}={\cal C}\hat\rho_{\rm in}$. In this case, the
coherent information is defined as
\begin{equation}\label{Ic} I_c=S[\hat\rho_{\rm out}]-S_e.
\end{equation}

\noindent Here, $S[\hat\rho]=-{\rm Tr}\,\hat\rho\log_2\hat\rho$ is the von
Neumann entropy, which generalizes the classical Boltzmann-Shannon entropy
$S[P(x)]=-\sum P(x_i)\log_2P(x_i)$ of a random variable $x=\{x_1,x_2,
\ldots\}$ with the probability distribution $P(x)=\{P(x_1),P(x_2),\dots\}$;
$S_e$ is the so-called entropy exchange with the reservoir responsible for
the noise in the channel. This definition is a direct generalization of
classical Shannon's information measure \cite{8} to the channels whose both
inputs and outputs are quantum.

In the case of the error-free transmission of all $M$ possible values of a
quantity  $x$, the classical Shannon information is given by $I=\log_2 M$.
For this choice of the logarithm base, a ``bit'' is conventionally
introduced as a unit of measure. If the transmitted values of $x$ have
unequal probabilities described by the probability distribution $P(x)$, the
above definition applies not directly to $x$, but rather to an ergodic
sequence $x_k,$ $k=1,\dots,n$ of statistically independent copies of the
variable $x$ with the probability distribution $P(x_1)\cdot\dots \cdot
P(x_n)$. In this case, as $n\to\infty$, the set of sequences with
asymptotically nonzero probabilities consists of $M_n =2^{S(P)}$
approximately equally probable members; the information per letter is
therefore $(\log_2 M_n)/n=S(p)$. This result, which, in particular, plays a
fundamental role in statistical physics, makes it possible to ascribe the
amount of information $I=S(P)$ to the noiseless transmission of all possible
values of the variable x with the probability distribution $P(x)$. If errors
may occur during the transmission, the corresponding nontrivial
information-transmission channel is described by the conditional probability
distribution $P(y|x)$ of the output able y for a fixed input x. In this
case, for long ergodic sequences, the accurately transmitted information per
letter is given by Shannon's {\em mutual information}
\begin{equation}\label{shannon}
I=S(P_x)+S(P_y)-S(P_{xy})=S(P_y)-\sum\limits_x S[P(y|x)]P(x).
\end{equation}

\noindent Here, $P_x$, $P_y$ and $P_{xy}$ denote respectively the
probability distributions for the input $x$, the output $y$, and the pair
$x,y$. The first equality in (2) demonstrates the symmetric (mutual) nature
of the Shannon information with respect to the input and the output. The
second equality represents the information as the difference between the
entropy of the output variable $y$ and the average of the variable $y$
introduced by the channel during the transmission of a given letter $x$. The
meaning of the last equality is most evident for the channel, in which the
transmitted letters $x$ are mapped to lapping subsets $M_x$ of the set of
possible values of $y\in\bigcup M_x$, i. e., the distortions are reduced to
the scatter of output variable $y$ within the regions $M_x$. The transmitted
information is defined in this case as the difference between the full
entropy of the variable $y$ and the average entropy of the subsets $M_x$.

The basic definition of the {\em coherent information} is given by the
relationship $I_c=\log_2\dim H$ where $H$ is the Hilbert space of the input
quantum system all the states of which are transmitted without errors. The
only name for the unit of quantum information is provided the term
``qubit,'' which is used in quantum computing theory and which corresponds
to a two-level quantum system with the dimensionality $\dim\,H{=}2$. The
essentially new element of this theory is the quantum nature of the
transmitted data, which can be found in any coherent superposition of basis
elements. Using the same line of reasoning as in the previous paragraph, one
can show that, if the distribution of the input states is described by the
density matrix $\hat\rho_{\rm in}$ and the channel transmits quantum states
$\psi\in H$ noiselessly, a measure of quantum information is provided by the
von Neumann entropy, i. e., the direct operator generalization of the
expression for the classical entropy. The simplest channel realizing
noiseless information transmission is, for example, the dynamic quantum
evolution of a closed system between two time instants, $t = 0$ and $t
\geqslant 0$.

If the quantum channel is noisy, the output state is a linear transform of
the input state: $\hat\rho_{\rm out}= {\cal C}\hat\rho_{\rm in}$. The role
of the channel superoperator ${\cal C}$ is then analogous to that of the
aforementioned conditional probability distribution $P( y|x)$ of a classical
channel. The quantum generalization of Shannon's definition (2) is based on
the expression behind the second equality, the first term of which---the
quantum entropy of the output---has a unique quantum generalization in the
form of the von Neumann entropy. The second term, which describes the
entropy introduced by the channel (the so-called entropy exchange $S_e$),
should be zero for the error-free transmission, i. e., for the identity
superoperator ${\cal C}={\cal I}$, in the quantum case as well. On the other
hand, if the input state is pure (the analog of a determinate classical
state), this term must coincide with the output entropy, which in this case
is completely due to the channel. These requirements can be satisfied if,
instead of the input quantum system, one considers its extension $H\otimes
H'$, where the degrees of freedom of $H'$ do not interact with those of the
channel, and the state $\hat\rho^{}_P$ of the joint system is pure and such
that its averaging produces the original state $\hat\rho_{\rm in}$ \cite{3}.
This procedure of replacing the initial quantum system is termed the
purification of a mixed quantum state. The corresponding transformation that
the channel performs on the extended quantum system is ${\cal C}\otimes{\cal
I}$, where ${\cal I}$ ensures that the variables of the ancillary system
remain unchanged. The resulting entropy exchange then coincides with the
entropy of the transformed extended system. The explicit form of the
purified state in $H\otimes H$ (i. e., in the case $H'=H$) is implicitly
contained in an expression derived in \cite{4}, which yields
\begin{equation}\label{rhoP}
\hat\rho^{}_P=\sum\limits_{ij}\sqrt{p_ip_j}\ket{i}\bra{j}\otimes \ket{i^*}
\bra{j^*},
\end{equation}

\noindent where $p_i$, $\ket{i}$, and $\bra{j}$ are respectively the
eigenvalues, and the right and left eigenvectors of the density matrix
$\hat\rho_{\rm in}$ and $\ket{i^*}$ ¨ $\bra{j^*}$ stand for the complex-
conjugate vectors. The purified state is thus combined from the input state
and its ``mirror reflection'' \cite{9}. The corresponding entropy exchange
is
\begin{equation} \label{Se} S_e=S(\hat\rho_\alpha),
\end{equation}

\noindent where
\begin{equation}\label{rhoalpha0}
\hat\rho_\alpha=({\cal C}\otimes{\cal I})\hat\rho^{}_P.
\end{equation}

\noindent In the general case, the channel transformation ${\cal C}$ can
describe the information transfer to an output system with a nonidentical
Hilbert space, $H_{\rm out}\neq H$.

In view of possible physical applications, we consider it important to give
an adequate physical interpretation of the density matrix (5) introduced in
\cite{4}, as well as the density matrix of the purified state (3) introduced
here. Both of these definitions are based on the aforementioned mathematical
arguments. Expression (3) describes the joint state of the input--mirror
reflection system, which gives rise to the input--output quantum-mechanical
state after the transmission. In the classical theory, the state (5)
corresponds to the conditional distribution $P(y|x)$ of the output for a
fixed input and, at the same time, to the averaging over the input states in
accordance with the distribution $P(x)$. The conditional distribution is
represented by the superoperator ${\cal C}$, whereas the averaging over the
input is represented by the structure of the wave function $\Psi^{}_P=\sum
\sqrt{p_i}\ket{i}\ket{i^*}$ that corresponds to the purified state (3). This
two-particle state is entangled, that is, it cannot be decomposed as a
statistical mixture of density matrices of the type $\ket{\psi_i}
\ket{\varphi_i}\bra{\varphi_i} \bra{\psi_i}$, which correspond to pure
direct products $\ket{\psi_i} \ket{\varphi_i}$ of one-particle states. Its
purely quantum fluctuations reproduce the mixed-nature fluctuations of the
density matrix $\hat\rho_{\rm in}$, which is defined in the first of the
subspaces involved in the direct product $H\otimes H$. Thus, the density
matrix (5) describes the state of the input--output system where the input
is replaced by its mirror reflection \cite{9}. It defines the entropy
exchange of the channel and, according to its physical meaning, is
qualitatively different from the usual one-instance density matrix since the
corresponding nonzero entropy appears as a result of the transformation of
the input state during its transmission in the channel. Unlike the usual
two-particle density matrix, this matrix always corresponds to a pure state
and zero entropy if the channel is noiseless.

\section{The model of the considered channel}\label{section:model}

The total dynamic system involved in the considered problem consists of a
$\Lambda$-system, which, in turn, is constituted by three levels
$\ket{1},\ket{2}$ and $\ket{3}$, and the field of free photons described by
the Hilbert space $H_F$. The corresponding total Hilbert space has the form
$H_\Lambda \otimes H_F$, where $H_\Lambda$ denotes the three-dimensional
Hilbert space of the $\Lambda$-system. For the purposes of the most reliable
storage and efficient manipulation of quantum information, the most
interesting systems are those where the transitions
$\ket{1}\leftrightarrow\ket{3}$ and $\ket{2}\leftrightarrow\ket{3}$ are
allowed and the transition $\ket{1}\leftrightarrow \ket{2}$ is forbidden.
The set of radiatively stable states $\ket{1}$ and $\ket{2}$ can then be
viewed as a {\em ground} two-level state (GTLS) of a $\Lambda$-system, which
a qubit of quantum information. To extract this information in the form of
photon field excitation, two laser fields, $E_L \cos(\omega_Lt+\varphi_1)$
and $E_L' \cos(\omega_L't+\varphi_2)$, are used, with their optical
frequencies $\omega_L$ and $\omega_L'$ satisfying the Raman resonance
condition $\delta_R\approx0$ \cite{6}, where
\begin{equation}\label{raman}
\delta_R=\omega_L'-\omega_L-\omega_{12}
\end{equation}

\noindent is the Raman detuning. The amplitudes $E_L$ and $E_L'$, the phases
$\varphi_{1,2}$, and the relaxation parameters of the $\Lambda$-system
determine the response of the $\Lambda$-system and the state of the photon
field at any time instant $t > 0$.

The study of coherent information extracted in this way simultaneously
answers the fundamental question of whether quantum information can be
extracted using a classical excitation, which, if weak, can only lead to the
quasi-classical excitation of the photon field to a coherent state. As far
as the TLA is concerned, this question should generally be answered in the
positive way, as it follows, for example, from the results of \cite{10}.
This is due to the essentially quantum character of an atom, which can be
viewed as a converter of the classical laser field.

In its pure form, the problem of the $\Lambda$-system photon field channel
can be considered only in the case of its single utilization since a reset
of the channel involves the introduction of new quantum systems
(reservoirs), which bring in new quantum information. A single extraction of
quantum information is realized by a pulsed laser excitation, which
entangles the initial state of the atom with the states of the photon field.
Apart from the obvious qualitative differences between a $\Lambda$-system
and a TLA, this process is analogous to the atom--field channel considered
in \cite{5}. Thus, the initial analysis of this problem naturally leads us
to the problem of the pulsed excitation that should be applied at times
shorter than the radiative lifetime to avoid decay-induced distortions and
thereby provide the maximum flexibility of the control.

\subsection*{3.1. Calculation of the coherent information in the case of
pulsed excitation} \label{subsection:pulse}

In the case of pulsed excitation, the information is transferred from a
specified initial state to a state of the photon field created due to the
action of exciting laser pulses on the $\Lambda$-system. The input of the
channel is the state at the time instant $t = 0$, specified by an arbitrary
$2\times2$ density matrix of the form
\begin{equation}\label{rhoin}
\hat\rho_{\rm in}= \sum\rho_{kl}\ket{k}\bra{l}
\end{equation}

\noindent where the matrix $(\rho_{kl})$, $k,l=1,2$ is positive-definite and
the trace is equal to unity, $\rho_{11}{+}\rho_{22}=1$. The initial state of
the $\Lambda$-system is then given by the same matrix (7) viewed as a
three-level system operator (for the considered class of initial states, we
assume that the population of the excited state $\ket3$ is zero). In the
rotating-wave approximation, the Liouvillian ${\cal L}$ does not contain any
relaxation parameters if we restrict ourselves to laser pulses with
durations $\tau_p\ll\gamma^{-1}$, where $\gamma$ is the rate of radiative
decay. Consequently, any pure initial state $\psi=c_1\ket{1}+c_2\ket{2}$ is
transformed to a pure excited state of the form
\begin{equation}\label{psi}
\psi(0)=\left(\begin{array}{c}
  \psi_1(0) \\
  \psi_2(0) \\
  \psi_3(0)
\end{array}\right)=\sum\limits_{k=1}^3 (U c)_k\ket{k},\quad
c=\left(\begin{array}{c}
  c_1 \\
  c_2
\end{array}\right),\quad
U=\left(\begin{array}{cc}
  U_{11} & U_{12} \\
  U_{21} & U_{22} \\
  U_{31} & U_{32}
\end{array}\right).
\end{equation}

\noindent This state subsequently relaxes by emitting photons and becomes a
metastable state of the form $\hat\rho_f=\sum_{k,l =1}^2
\rho_{fkl}\ket{k}\bra{l}$. The latter state can also contain some about the
initial state, making the channel $\hat\rho_{\rm in}\to \hat\rho_f$ another
subject for consideration. However, the optimization of this channel over
the exciting is trivial: the optimum corresponds to no excitation. The
nontrivial channel that is considered here involves the transfer of
information to the photon field.

Photons are emitted exclusively by the excited state $\ket{3}$. If the
interaction Hamiltonian preserves the total number of excitation quanta in
the atom + field system, the dynamics can be solved exactly: it is reduced
to the emission of a superposition of two photons corresponding to the
transitions $\ket{3}\to\ket{1}$ and $\ket{3}\to \ket{2}$. The photons are
described in the same fashion as in the case of the TLA except for the fact
that one has to take into account the connection via the common excited
level $\ket{3}$. In contrast to the TLA, the photon emitted by the
$\Lambda$-system splits into a superposition of the two photons that
correspond to its two transitions and oscillate at the respective
frequencies $\omega_{13}$ and $\omega_{23}$.

Thus, at any time instant $t\gtrsim\gamma^{-1},\tau_p$, the state of the
atom + field system is described by the wave function
\begin{eqnarray}\label{photons}
\Psi(t)&= &\displaystyle\psi_1(0)e^{-i\omega_1t}
\ket1\ket0+\psi_2(0)e^{-i\omega_2t}\ket2\ket0+ \psi_3(0)
e^{-i(\omega_3+\Lambda)t-\gamma t/2}\ket3\ket0+\nonumber\\ &
&\vphantom{\int} \displaystyle \psi_3(0) \sqrt{1-e^{-\gamma t}}
\left(\alpha_1\ket{1}\ket{\psi_{13}}+ \alpha_2\ket2\ket{\psi_{23}}\right).
\end{eqnarray}

\noindent Here, $\omega_1$, $\omega_2$ and $\omega_3$ are the
eigenfrequencies of the energy levels; $\Lambda_{13}$ and $\Lambda_{23}$ are
the radiative frequency shifts due to the transitions 1$\leftrightarrow$3
and 2$\leftrightarrow$3; $\gamma_{13}$ and $\gamma_{23}$ are the rates of
the radiative decay from level 3 to levels 1 and 2, respectively;
$\gamma=\gamma_{13}+ \gamma^{}_{23}$ is the total decay rate of the excited
state 3; $\Lambda= \Lambda_{13}+ \Lambda_{23}$ is the total frequency shift;
$\alpha_1=\gamma_{13}/\gamma$ and $\alpha_2=\gamma_{23}/\gamma$ describe the
distribution the radiative decay over the two considered dipole transitions;
$\ket{0}$, $\ket{\psi_{13}}$ and $\ket{\psi_{23}}$ are, respectively, the
vacuum state of the photon field and the states with a single photon at the
frequency $\omega_{13}$ and $\omega_{23}$. The first three terms describe a
superposition of the atomic states in the absence of photons; the last term,
a superposition of two photon states that are entangled with the
corresponding levels 1 and 2 of the $\Lambda$-system. We ignore the
decoherence in this approach as the photon wave functions $\ket{\psi_{13}}$
and $\ket{\psi_{23}}$ are assumed to be known exactly due to the fact that
we impose no restrictions on the way the photon signal can be used
\cite{11}. As opposed to the TLA--photon field channel, where the relevant
set of photon states constitutes a two- level system, here it consists of
three states, $\ket{0}$, $\ket{\psi_{13}}$ and $\ket{\psi_{23}}$, which are
viewed, similar to \cite{5}, as the basis states, in terms of which the
subsequent analysis of the channel transformation is performed. It is easy
to verify that the partial density matrices of the atom and the field that
are associated to the wave function (9) correspond to the description of the
radiative decay in terms of the relaxation dynamics of open systems. As for
the fast-oscillating exponents before the atomic wave functions
$\ket{1,2,3}$ one can easily see that, being associated to wave functions,
they do not affect the information characteristic and therefore can be
neglected during their calculation.

The function (9) describes the isometric transformation $V$: atom
$\rightarrow$ atom+field, i. e., the transformation $\psi(0)\to\Psi(t)$ of
the form $\Psi(t)=\sum_{k\alpha l}V_{k\alpha,l}\psi_l(0)\ket{k}
\ket{\alpha}$, where $k,l=1,2,3$ number the atomic states, and $\alpha$
numbers the states of the photon field. Expressing $\psi_l(0)$ through the
initial state of the $\Lambda$-system before the application of the laser
pulses, we obtain an isometric mapping of the GTLS to a state of the atom +
field system described by the matrix $W_{k\alpha,m}=\sum_l V_{k\alpha,l}
U_{lm}$. The corresponding superoperator of the GTLS--photon field channel
has the form ${\cal C}=\sum\hat s_{mn}\bra{m} \odot \ket{n}$ (\cite{5},
where $\odot$ is the substitution symbol that should be replaced by the
density matrix to be transformed, $\hat\rho_{\rm in}$, and $\hat s_{mn}$ is
the 2$\times$2 matrix of the photon-field operators, which are represented
by 3$\times$3 matrices with the matrix elements
\begin{equation}\label{smn}
(\hat s_{mn})_{ \alpha\beta}=\sum\limits_{k=1}^3
W_{k\alpha,m}^{}W_{k\beta,n}^*.
\end{equation}

\noindent Here, the summation is performed over all three atomic states, and
the output indices number the photon states 0, $\psi_{13}$ and $\psi_{23}$.

If the $\Lambda$-system is excited by two rectangular pulses with a
negligible detuning and a duration $\tau_p$, the matrix $U$ defining the
initial atomic state depends on three parameters: $\theta=\sqrt{\Omega_1^2 +
\Omega_2^2}\tau_p$ is the total action angle of the two pulses; $\chi =
\arctan(\Omega_1/ \Omega_2)$ is the angle describing the distribution of the
field intensity over the pulses; $\varphi$ is the relative phase of the
pulses. By varying these parameters we can optimize the amount of
information that is transferred from the initial state to the states of the
photon field. In the case of the maximum-entropy initial state $\hat
\rho_{\rm in}=\hat I/2$, which provides the maximum amount of information
for a certain choice of the laser parameters (one can show this using the
symmetry arguments), the coherent information does not depend on the
parameters $\chi$, $\varphi$, which can therefore be neglected. The
corresponding set of operators $\hat s_{mn}$ has the form
\begin{equation}\label{smn0}
\small
\begin{array}{c}
\hat s_{11}=\left(\begin{array}{ccc}
  \cos^2\frac{\theta}{2}+e^{-\gamma t}\sin^2\frac{\theta}{2}&
  -\frac{i}{2}\sqrt{\alpha_1}\sqrt{1-e^{-\gamma t}}\sin\theta& 0\\
  \frac{i}{2}\sqrt{\alpha_1}\sqrt{1-e^{-\gamma t}}&\alpha_1
  (1-e^{-\gamma t})\sin^2\frac{\theta}{2}& 0 \\
  0 & 0 &\alpha_2(1-e^{-\gamma t})\sin^2\frac{\theta}{2}
\end{array}\right),\\
\hat s_{12}=\left(\begin{array}{ccc} 0&0&0\\ 0&0&0\\
i\sqrt{\alpha_2}\sqrt{1-e^{-\gamma t}}\sin\frac{\theta}{2}&0&0
\end{array}\right),\;
\hat s_{21}=\left(\begin{array}{ccc} 0&0&-i\sqrt{\alpha_2}\sqrt{1-e^{-\gamma
t}}\sin\frac{\theta}{2}\\ 0&0&0\\ 0&0&0
\end{array}\right),\;
\hat s_{22}=\left(\begin{array}{ccc} 1&0&0\\ 0&0&0\\ 0&0&0
\end{array}\right).
\end{array}
\normalsize
\end{equation}

After defining the operators $\hat s_{mn}$, which determine the
superoperator $\cal C$ of the considered channel, we carry out the rest of
the calculation using the general formulas given in \cite{5}.

\section{Results of calculation}
\label{section:results}

\begin{figure}
\begin{center}
\epsfysize=5.cm \epsfclipon \leavevmode \epsffile{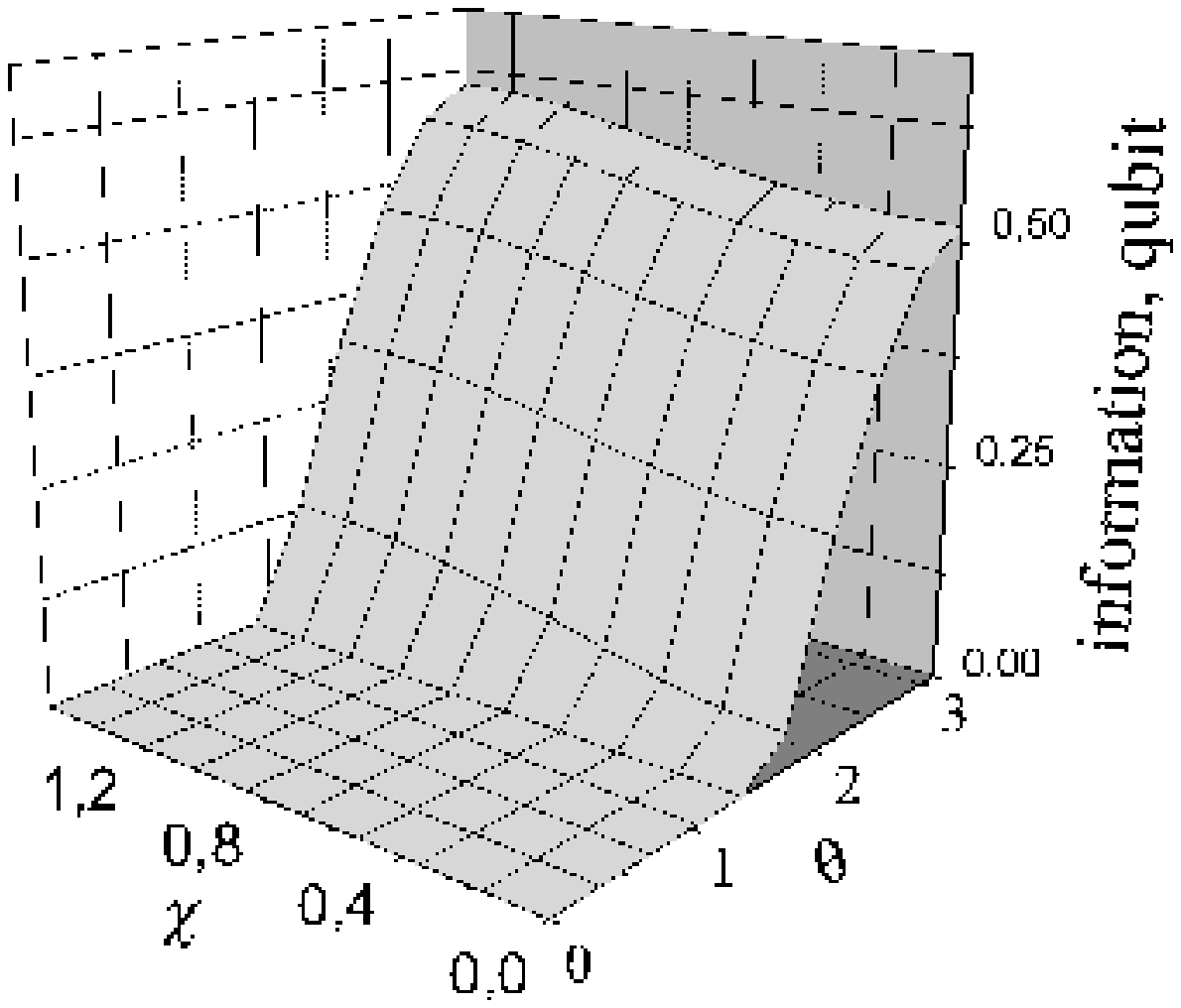} \epsfysize=5.cm
\epsfclipon \leavevmode \epsffile{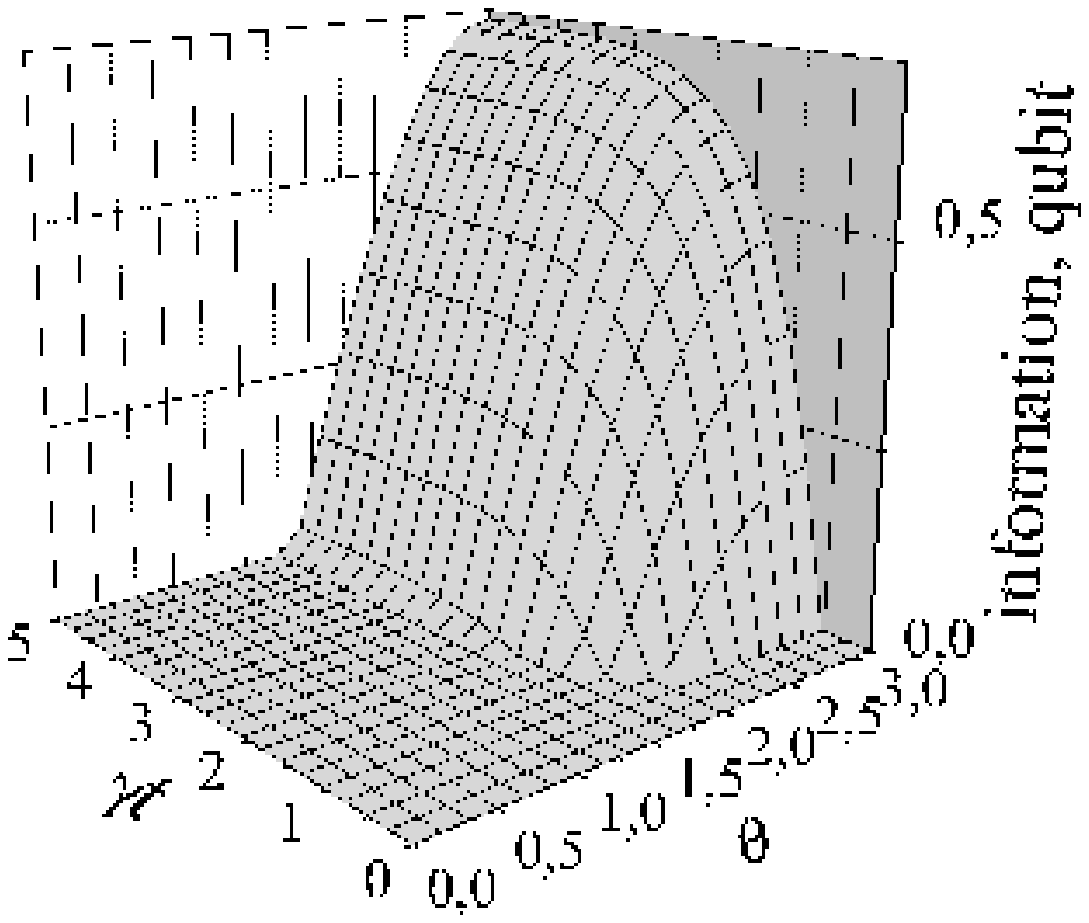}
\end{center}
\caption{Coherent information in a symmetric $\Lambda$-system (a) as a
function of the total action integral $\theta=\Omega\tau_p$ and the angle
$\chi$ describing the distribution of the driving field amplitudes over the
levels 1 and 2 with $t\to\infty$ (the input density matrix is diagonal with
the matrix element $\rho_{11}=1/4$) and (b) as a function of the total
action integral $\theta=\Omega \tau_p$ and the dimensionless time $\gamma t$
(the input density matrix is the maximum-entropy state $\hat I/2$).}
\label{fig1}
\end{figure}

\begin{figure}
\begin{center}
\epsfysize=5.cm \epsfclipon \leavevmode \epsffile{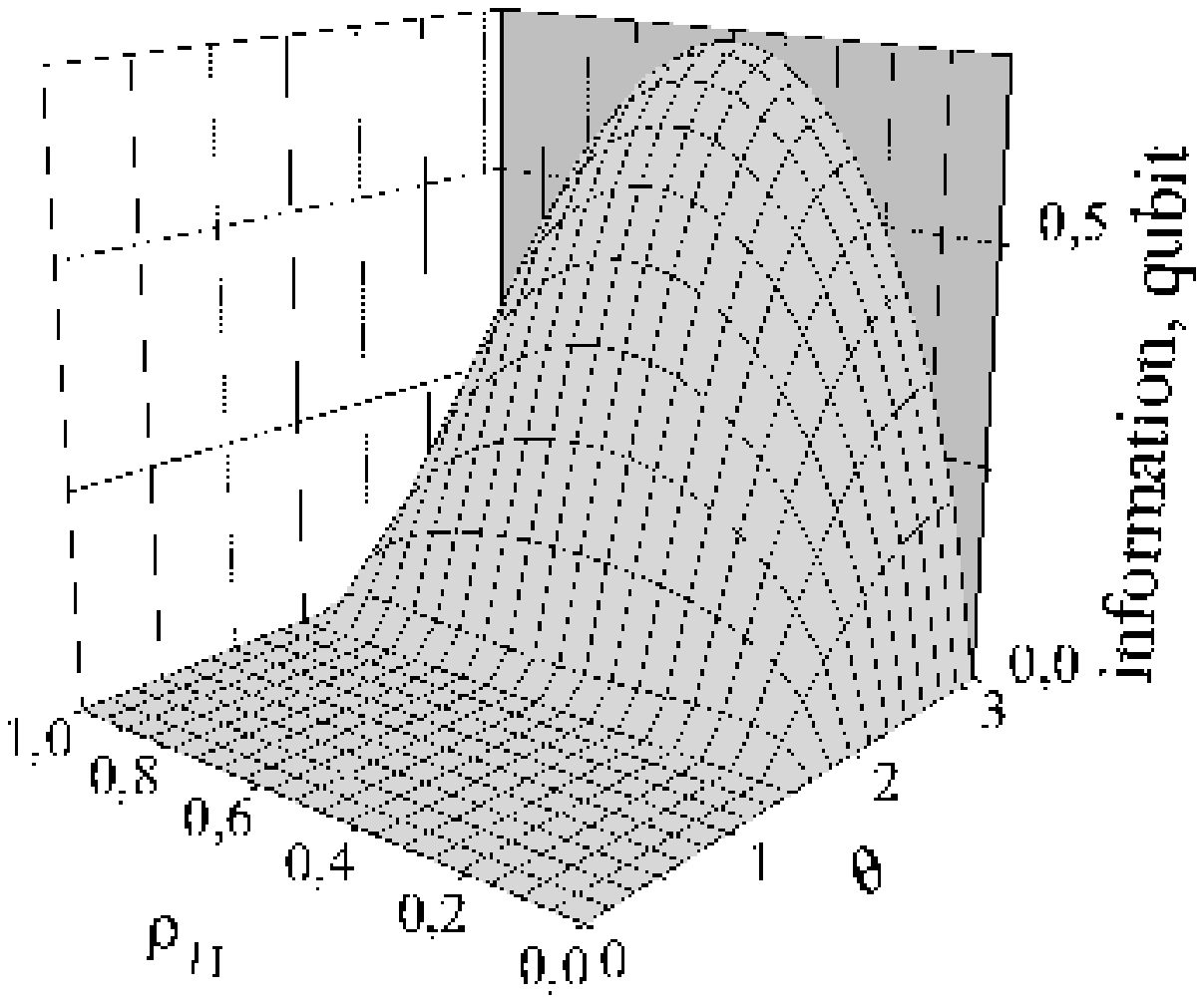} \epsfysize=5.cm
\epsfclipon \leavevmode \epsffile{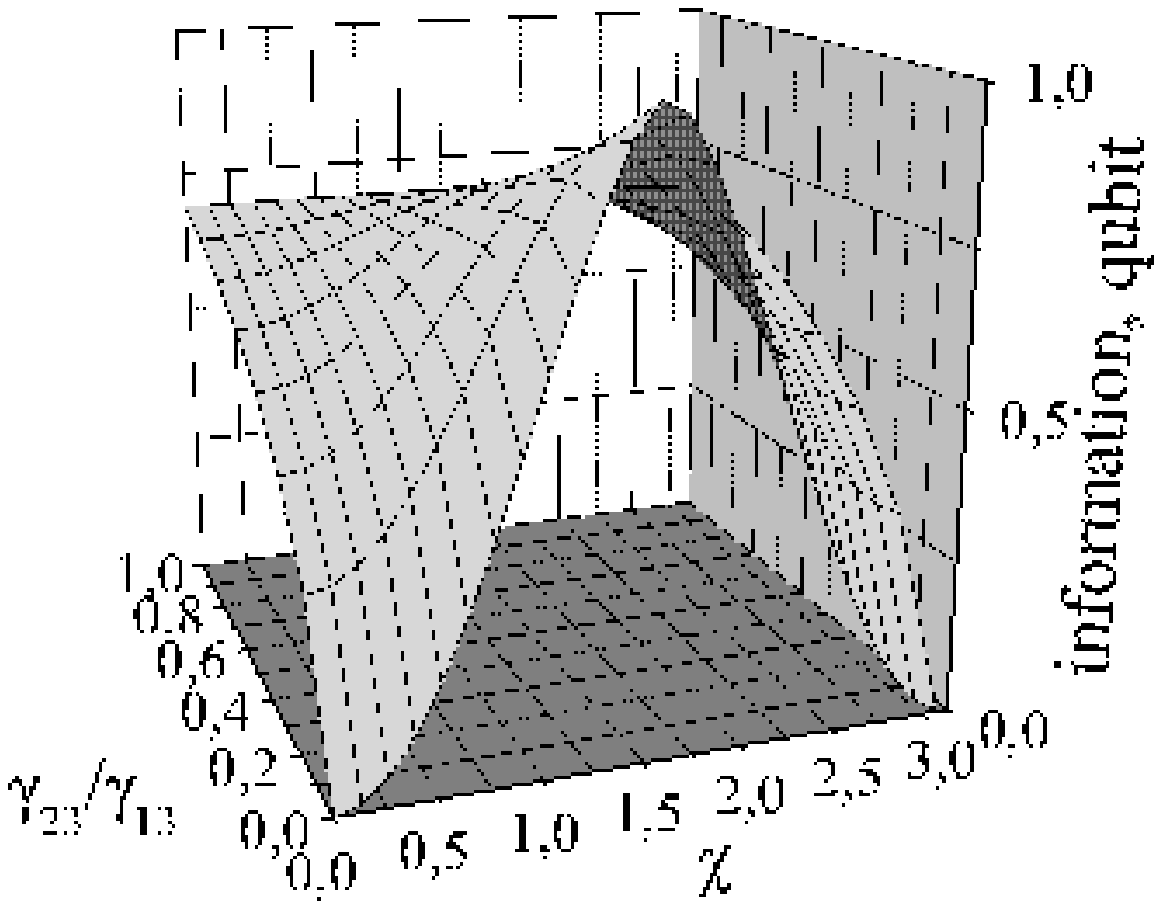}
\end{center}
\caption{(a) Coherent information in a symmetric $\Lambda$-system as a
function of the total action integral $\theta=\Omega\tau_p$ and the matrix
element $\rho_{11}$ of the diagonal density matrix with $t\to\infty$. (b)
Coherent information as a function of the system asymmetry $\gamma_{23}/
\gamma_{13}$ and the driving field distribution angle $\chi$ with
$t\to\infty$ for $\hat \rho_{\rm in}=\hat I/2$.}\label{fig2}
\end{figure}

Figures \ref{fig1} and \ref{fig2}  show the dependence of the coherent
information in a symmetric $\Lambda$-system ($\gamma_{13}=\gamma_{23}$) on
the relevant parameters of the considered channel. The total action integral
$\theta=\Omega\tau_p$ is one of such parameters. In the case of a symmetric
system and the initial state $\hat\rho_{\rm in}=\hat I/2$, the coherent
information does not depend on the distribution of the driving intensity
over the two laser fields. Input density matrices of the general form do not
provide the maximum amount of information; moreover, in this case, the
coherent information depends on the intensity distribution of the exciting
pulses, i. e., the parameter $\chi$. Figure 1a shows the coherent
information at $t\to\infty$ as a function of the angle $\chi$ in the case of
the diagonal input density matrix $\hat\rho_{\rm in}=\left(\begin{array}{cc}
  1/4 & 0 \\
  0 & 3/4
\end{array}\right)$.
Figure 1b shows the coherent information as a function of the dimensionless
time $\gamma t$ in the case of the maximum-entropy input density matrix
$\hat\rho_{\rm in}=\hat I/2$. Figure 2 shows the coherent information of the
symmetric system at $t\to\infty$ as a function of the input density matrix
$\hat \rho_{\rm in}$.

It follows from these results that, in the case of a symmetric
$\Lambda$-system, the maximum of the coherent information is achieved with
the following parameters: $$\hat\rho_{\rm in}=\frac{\hat I}{2},\quad
t\to\infty,\quad \theta=\pi.$$

\noindent The first condition ensures that the amount of quantum information
at the input is maximum; the second, that it is maximally transferred to the
photon field; and the third, that the population from the ground state is
maximally transferred to the excited one. The corresponding maximum amount
of quantum information, i. e., the potentially attainable information
capacity of the channel, is $I_c=0,6887$.

Figure 2b shows the dependence of $I_c$ on the asymmetry degree $\gamma_{23}
/ \gamma_{13}$ and the driving-intensity distribution angle $\chi$ in the
case of a fully emitted photon, i. e., $\gamma t\to\infty$, the optimal
value of $\theta=\pi$, and $\hat\rho_{\rm in}=\hat I/2$. The maximum,
$I_c=1$ qubit, is reached at $\chi=\pi/2$ and $\gamma_{23}/\gamma_{13}=0$,
which corresponds to the reduction of the $\Lambda$-system to a two-level
system. The optimization over the relative phase of the pulses $\phi$ is
redundant for any values of the other parameters since this corresponds to a
variation in the coding method but not the amount of information.

\section{Conclusions}\label{section:conclusion}

Thus, the utilization of a symmetric $\Lambda$-system for the transfer of
information to the photon field somewhat reduces the information capacity of
the channel with respect to the emission of a two-level system. This
deficiency in the information capacity can be viewed as the commission for
the radiative stability of the stored qubit and the advantages in the
radiative manipulation of its states. Therefore, the photon field cannot,
even in principle, be used for a quantum computation as an equivalent of the
information qubit stored in the ground state of the $\Lambda$-system.
Nevertheless, as far as multiqubit operations are concerned, the information
loss amounts to only 31\%, which demonstrates the possibilities for a fairly
efficient utilization of the photon field for the physical conversion of
coherent quantum information. The performed analysis also provides a general
idea about the information losses intrinsically related to the extraction of
the information about the ground state of the $\Lambda$-system by the
methods of laser spectroscopy.

\section*{Acknowledgments}

This research was supported by the ``Nanotechnologies'' and ``Fundamental
Metrology'' programs of the Ministry of Science and Technologies of Russian
Federation.


\begin{thebibliography}{99}

\bibitem[]{}\section*{References\hphantom{AA}}

\bibitem{1} Williams, C. P. and Clearwater, S. H., 1998, {\em Explorations
in Quantum Computing} (New York: Telos/ Springer-Verlag).

\bibitem{2}Preskill, J., 1999, {\em Lecture Notes on Physics 229: Quantum
Information and Computation}, http://www.
theory.caltech.edu/people/preskill/ph229/.

\bibitem{3}Barnum, H., Nielsen, M. A., and Schumacher, B., 1998, {\em Phys.
Rev. A}, {\bf 57}, 4153.

\bibitem{4}Lloyd, S., 1997, {\em Phys. Rev. A}, {\bf 55}, 1613.
\bibitem{5}Grishanin, B. A. and Zadkov, V. N., 2000, {\em Phys. Rev. A}, {\bf
62}, pp.~032303--12.

\bibitem{6}Arimondo, E., 1996, {\em Progress in Optics}, Wolf, E., Ed.
(Amsterdam: Elsevier), vol. 35, p. 257.

\bibitem{7}Bargatin, I. V., Grishanin, B. A., and Zadkov, V. N., 2000,
{\em Phys. Rev. A}, {\bf 61}, 052305.

\bibitem{8}Gallagher, R. G., 1968, {\em Information Theory and Reliable
Communication} (New York: John Wiley and Sons).

\bibitem{9}The complex conjugation, which is introduced here and was
absent in \cite{4}, is required for the invariance of the considered
representation with respect to rotations in the subspaces corresponding to
degenerate eigenvalues of the density matrix. For real density matrices
$\hat\rho_{\rm in}$ with nondegenerate spectra, this refinement is
unnecessary.

\bibitem{10}van Enk, S. J. and Kimble, H. J., 1999, quant-ph/9908082.

\bibitem{11}To take into account the restrictions due to the
actual geometry of the photodetector and its finite time resolution, one
should introduce a corresponding transfer function. Obviously, it is here
that lies the border between the description of the atom + field system
dynamics in terms of a unitary transformation and a relaxation random
process.

\end{thebibliography}
\end{document}